\documentstyle[11pt,newpasp,twoside]{article}
\def\edcomment#1{\iffalse\marginpar{\raggedright\sl#1\/}\else\relax\fi}
\reversemarginpar

\begin{document}
\title{The Shape of the Big Blue Bump as Revealed by Spectropolarimetry}
\author{R. Antonucci}
\affil{University of California, Santa Barbara, Department of Physics, Santa Barbara, CA  93106-9530}
\author{M. Kishimoto}
\affil{University of Edinburgh, Institute for Astronomy, Royal
Observatory, Blackford Hill, Edinburgh EH9 3HJ, UK}
\author{C. Boisson}
\affil{Observatoire de Paris-Meudon, DAEC, 92195, Meudon, France}
\author{O. Blaes}
\affil{University of California, Santa Barbara, Department of
Physics, Santa Barbara, CA  93106-9530}

\begin{abstract}

Thermal models for the quasar Big Blue Bump generally lead to bound-free
continuum features, which may be in absorption or emission.  Searches
for Lyman edges attributable to quasar atmospheres (in particular
accretion disk atmospheres) have been ambiguous at best, but
various relativistic and non-LTE effects may make them hard to detect.

The Balmer edge features tentatively predicted by such models
might be easier to detect since they'd form farther out in the
potential wells.  These can be sought in certain cases using
spectropolarimetry to remove the effects of atomic emission (i.e.
the Small Blue Bump) from the spectrum. We do find apparent Balmer
edges in absorption in several quasars using this method! Although
the features we see are believable and apparently common, more
data and more modeling are needed to verify that they are best
interpreted as atmospheric Balmer edges.

\end{abstract}

\section{Introduction}

Early models of quasar accretion disks\footnote{It must be
emphasized that almost all published accretion disk models are
quasi-static and based on the Shakura-Sunyaev model.  This is
known to be qualitatively and quantitatively inconsistent with the
observations of variability, spectral energy distributions and
other properties.  A sampling of references for this statement is
the following:  Alloin et al 1985, Antonucci 1988, 1999; Krolik et
al 1991;  Gaskell et al 2003; D vanden Berg 2003, these
proceedings; O Blaes 2003, these proceedings. However, the models
may indicate in a general way the behavior of optically thick
thermal radiation in the case of energy deposition at large
optical depth, especially if the emission follows the potential
drop as in the Shakura-Sunyaev paradigm.}predicted strong Lyman
edges in absorption, as expected for stars of moderately high
temperature, due to source function gradients in the atmospheres
(e.g. Kolykhalov and Sunyaev 1984).  Any such edges aren't easily
seen (Antonucci, Kinney and Ford 1989 and later
papers)!\footnote{Also note that partial Lyman Limit Systems such
as those of S5 0014+81 can give false positives for atmospheric
edges (e.g. Fig 1a of Antonucci et al 1989 and the accompanying
discussion). Thus no edges should be claimed to be atmospheric
unless it has been shown that they lack accompanying narrow
absorption lines. Candidates for broadened edges have generally
failed this test, when the test has been made (Koratkar, Kinney
and Bohlin (1992)).}

Changes in spectral slope do appear to be common in the vicinity
of the Lyman limit (Zheng et al. 1995 and later papers).  Such
spectral breaks might be consistent with smeared out and/or
Comptonized Lyman edge features (e.g. Lee, Kriss, \& Davidsen
1992, Hubeny et al. 2000).  However, detailed fits to the spectral
break observed in 3C~273 with the latest disk atmosphere models
still produce a local emission bump longward of the limit that is
not in agreement with the data (Blaes et al. 2001), and other
problem arise when fitting the reported spectral breaks as
atmospheric edges (Kishimoto et al, in prep).

Accretion disk models place the $\sim$4000A continuum around seven
times farther out in the gravitational potential well for peak
emissivity, compared with the Lyman edge region. This would
greatly reduce relativistic smearing. Also since Balmer continuum
absorption is not a resonance process, there is less chance of
being fooled by intervening absorbing matter.  In fact the
distinction between atmospheric and foreground absorption becomes
in part a semantic one in that case.

The Balmer edge behavior has never been discussed as a diagnostic of the
quasar continuum emission mechanism because very strong Balmer continuum
EMISSION and FeII features, attributed to the Broad Line Region,
contaminate this wavelength region.

Very fortunately there is a way around the problem of
contamination by atomic emission in some objects.  These have
slightly polarized BBBs, and UNPOLARIZED emission lines.  The
likely explanation is that there is some Thomson scattering
interior to the BLR, which polarizes the BBB only.\footnote{We
discuss the possibility that the observed polarization is due to a
synchroton component (Schmidt and Smith 2002) in Kishimoto et al
2003.   Here we just note that the existence of the edge feature
in polarized flux requires that at least some of the polarization
result form a thermal process.}

Thus a plot of polarized flux can cleanly shave off
the atomic emission.  Published data of moderate SNR confirm that
in these cases the Small Blue Bump is indeed unpolarized (e.g.
Antonucci 1988;  Schmidt and Smith 2000).  With recent
higher-SNR data from larger telescopes, we can determine the behavior
of the BBB alone at the Balmer edge postion.

\section{Observations and Results}

A total of around fifteen good moderate-redshift candidates for
polarized continuum and unpolarized atomic emission have been
taken from the literature.  We observed two of these objects with
the Keck I telescope on May 9, 2002 (UT), and reported the results
in Kishimoto et al 2003. The object with the better data, Ton 202,
clearly shows the polarized flux turning DOWN below 4000A, just
where the total flux turns UP because of the SBB.

Next we observed eleven quasars at the VLT Unit 3 (Melipal). These
objects show a rich variety of interesting behavior, and some are
unsuitable for our method because of substantial emission line
polarization.  For the suitable quasars, the polarized flux
spectra show absorption edges similar to that of Ton 202 in at
least two other cases (3C95 and 4C09.72)! The behavior thus seems
rather widespread.

A second run with Keck I on May 4, 2003 (UT) uncovered a similar
Balmer edge feature in two more objects. As noted by Schmidt and
Smith 2000, the polarization of individual objects varies over
time. This isn't surprising since the scatterers must reside
interior to or cospatial with the BLR.

\section{Discussion}
We often find that the polarized flux DECREASES at wavelengths
below ~4000A, strongly suggesting that the BBB has such a feature,
and is thus thermal emission from an optically thick source.
It may be that the total flux of the BBB has no feature at this
wavelength, but that instead the BBB percent polarization has the feature.
This type of behavior occurs in some models (Laor, Netzer and Piran
1990;  a rigorous calculation which gives rather different behavior
in some part of parameter space was later published as
Blaes and Agol 1996).  Still the implication is optically thick thermal
emission.

Although the observed Balmer edge behavior is qualitatively consistent
with thermal model predictions, more data and more modeling are needed
to ensure that this is the correct interpretation.

On a slightly different topic, a high point of the SDSS conference
for us was a lunchtime discussion of the polarization of 2MASS
quasars with Gary Schmidt.  His group has observed these objects
extensively (e.g. Smith et al 2003).  The objects were selected
for red J-K colors. Considering the nonzero redshifts, this means
in practice that they were selected for LARGE RATIOS OF HOT DUST
EMISSION TO BBB EMISSION. The polarization data are very
remarkable!  We'd have expected the 2MASS objects to be mostly
reddened quasars, with polarization due to dust scattering or
absorption.  Dusty far-IR selected quasars behave this way in
general (e.g. Hines et al 2001).  Dust scattering typically
produces very blue polarized flux spectra in the objects studied
in that paper and in other similar objects, at least in the
optical part of the spectrum.  (The shorter wavelengths are often
cut off by dust absorption.)  None of the 2MASS objects behave
that way. However many 2MASS quasars are highly polarized:  the
percent polarization is often $\sim10\%$, rather than $\sim1\%$ as
for our objects!  In general the broad lines ARE polarized in
objects selected this way, and so that this high polarization
arises on BLR scales of larger.  These properties make the
interpretion of the spectropolarimetry very different from that of
the low-polarization quasars described above.

Modest foreground reddening suppressing the BBB wouldn't necessarily produce
very red J-K colors since that wavelength interval is small and at relatively
long wavelength;  rather as noted
above it could indicate a large ratio of dust emission to BBB emission.  Actual
optical/NIR SEDs will help to distinguish the two possibilities.

Over the lunch able we considered the possibility that the BBB
emission is suppressed by foreground Thomson scattering.  The
(often very high) polarizations may be due to Thomson scattering
as well.  Large electron-scattering optical depths are very
reasonable for quasars and luminous Seyfert galaxies.  A powerful
argument is that the scatterers in objects like NGC1068 provide an
optical depth of 1$\%$ due to electron scattering (Miller,
Goodrich, and Mathews 1991; Antonucci et al 1994; Ogle et al 2003)
\footnote{This paper actually finds an angle-averaged optical
depth on $\sim$100pc scales of order a percent, but within the
ionization/scattering cone, a small covering factor of sight lines
with much higher Thomson depth.} and this polarization arises  on
$\sim$100pc scales (Antonucci et al 1994; Capetti et al 1995).
The outflowing scattering region is likely to be a wind which
originates near the sublimation (or ''ablation") radius of $<$1pc
(Krolik and Begelman 1986).  Thus given the density law for winds,
THE OPTICAL DEPTH OF THE WIND IS OF ORDER UNITY on pc scales.
(Ferland et al, this meeting, also argue for such
electron-scattering optical depths near the torus inner edge.)
Thus NGC1068 would look like a luminous Seyfert 1 or weak quasar
if seen from the polar regions, and would present a very
substantial electron-scattering optical depth to the BBB-emitting
region. Similarly many quasars show large X-ray absorption columns
with little or no optical reddening and absorption. (e.g. Salvati
and Maiolino 2000; Wilkes et al 2002).  These observations are
interpretable in the context of the scattering wind (Krolik and
Kriss 2001).

A delightful aspect of the 2MASS spectropolarimetric observations
is that the DIRECT, unpolarized line profile can often be seen
simultaneously with the Thomson-scattered line profile!  See for
example 2MASS 222202.2 +195231 in Fig 2 of Smith et al 2003. Thus
we can get an accurate temperature and outflow velocity, as well
as a good idea of the electon- scattering optical depth.   For the
object mentioned, the electron temperature comes out to one
million degrees (FWZI = 18,000 km/s) and the outflow velocity is
around 4000 km/sec (G. D. Schmidt, 2003 p.c.). However,
demonstration of large electron-scattering optical depth in front
of many quasar BBBs will entail many observational tests.

To summarize, the polarization property is underexploited in this field,
and it can provide many unique insights.

We thank Gary Schmidt and Pat Ogle for sharing their thoughts.  The work
was funded in part by NSF AST-0098719.

\end{document}